\documentclass[final,5p,times,twocolumn]{elsarticle}

\makeatletter
\def\ps@pprintTitle{%
 \let\@oddhead\@empty
 \let\@evenhead\@empty
 \def\@oddfoot{\centerline{\thepage}}%
 \let\@evenfoot\@oddfoot}
\makeatother

\usepackage{amssymb}
\usepackage{amsmath}

\usepackage{lineno}

\usepackage{siunitx}
\usepackage{float}

\usepackage{listings}
\usepackage{color}
\definecolor{green}{RGB}{0,150,0}
\definecolor{deepblue}{rgb}{0,0,0.5}
\definecolor{deepred}{rgb}{0.6,0,0}
\definecolor{deepgreen}{rgb}{0,0.5,0}
\newcommand{\Ecal}{\mathcal{E}}
\newcommand{\LINKTOCODE}{https://github.com/omelchert/py-fmas}
\newcommand{\LINKTOPAGES}{https://omelchert.github.io/py-fmas}

\usepackage[labelfont=bf,singlelinecheck=false]{caption}
\usepackage{hyperref}
\hypersetup{
    colorlinks,
    linkcolor=Blue
}

\newcounter{bla}

\journal{}

\begin{document}

\begin{frontmatter}



\title{A python package for ultrashort optical pulse propagation in terms of forward models for the analytic signal}

\author[add1,add2]{O. Melchert\corref{mycorrespondingauthor}}
\ead{melchert@iqo.uni-hannover.de}
\cortext[mycorrespondingauthor]{Corresponding author}

\author[add1,add2]{A. Demircan}
\ead{demircan@iqo.uni-hannover.de}

\address[add1]{Institute of Quantum Optics, Leibniz Universit\"at Hannover, Welfengarten 1, 30167 Hannover, Germany}
\address[add2]{Cluster of Excellence PhoenixD (Photonics, Optics, and Engineering - Innovation Across Disciplines), Welfengarten 1, Hannover, Germany}

\begin{abstract}
We present a flexible, open-source Python package for the accurate simulation
of the $z$-propagation dynamics of ultrashort optical pulses in nonlinear
waveguides, especially valid for few-cycle pulses and their interaction.
The simulation approach is based on unidirectional propagation equations for
the analytic signal. 
The provided software allows to account for dispersion, attenuation, 
four-wave mixing processes including, e.g., third-harmonic generation, and features various 
models for the Raman response.
The propagation equations are solved on a periodic temporal domain.
For $z$-propagation, a selection of pseudospectral methods is available.
Propagation scenarios for a custom propagation constant and initial field
pulses can either be specified in terms of a HDF5 based input file format 
or by direct implementation using a python script.
We demonstrate the functionality for a test-case for which an exact solution is
available, by reproducing exemplary results documented in the scientific
literature, and a complex propagation scenario involving multiple pulses.
The {\tt py-fmas} code, its reference manual, an extended user guide, and
further usage examples are available online at \url{\LINKTOCODE}.\\
\end{abstract}


\end{frontmatter}


\tableofcontents

\section{Introduction}
The accurate theoretical description of the propagation dynamics of ultrashort
optical pulses in nonlinear media requires flexible models, adaptable to a wide
variety of experimental conditions, and accurate approximation methods, 
facilitating the dynamical evolution of the optical field.
Typically considered models are, e.g., the forward Maxwell equation
\cite{Husakou:PRL:2001}, or nonlinear envelope equations
\cite{Brabec:PRL:1997}, such as the generalized nonlinear Schr\"odinger
equation \cite{Agrawal:BOOK:2013,Kivshar:BOOK:2003,Mitschke:BOOK:2010}.

Here, we introduce {\tt py-fmas}, a Python package for the accurate numerical
simulation of the complex propagation dynamics of ultrashort optical pulses in
nonlinear waveguides, especially valid for few-cycle pulses and propagation
scenarios involving multiple pulses with distinct frequencies.
The considered $z$-propagation models are formulated in terms of the
complex-valued analytic signal related to the real-valued optical field
\cite{Amiranashvili:PRA:2010,Amiranashvili:AOT:2011,Amiranashvili:BOOK:2016,Demircan:APB:2014,Demircan:OE:2014}.
Description of the field dynamics in terms of the analytic signal has several
advantages. For example, it allows to directly neglect non-resonant
contributions of four-wave mixing \cite{Amiranashvili:AOT:2011}, and to derive
models that are formally simpler than the forward Maxwell equation
\cite{Husakou:PRL:2001}. Still, additional nonlinear effects such as the Raman
effect can be included in a standard way
\cite{Demircan:OE:2014,Demircan:APB:2014}.
A further advantage of these models is that they are exempt from the slowly
varying envelope approximation (SVEA) commonly adopted for the derivation of
nonlinear Schr\"odinger type equations.
This limits the applicability of the latter models for the accurate simulation
of few-cycle pulses which do not satisfy the SVEA
\cite{Oughstun:PRL:1997,Oughstun:BOOK:2009}.
Nevertheless, as prominent limiting case, the envelope-based generalized
nonlinear Schr\"odinger equation
\cite{Agrawal:BOOK:2013,Kivshar:BOOK:2003,Mitschke:BOOK:2010}, including all
its usual effects, can be obtained from analytic signal based models
\cite{Amiranashvili:PRA:2010}.
For the dynamical evolution of the analytic signal, {\tt py-fmas} provides a
selection of propagation algorithms commonly used in nonlinear optics for the
solution of nonlinear Schr\"odinger type equations.

The remainder of the article is organized as follows. In
sect.~\ref{sec:prop_models} we introduce the considered propagation models, and
in sect.~\ref{sec:problem} we state the computational problem solved by the
provided software. In sect.~\ref{sec:algs} we detail the implemented numerical
methods. In sect.~\ref{sec:user_manual} we provide a brief user manual that
guides the reader through a workflow using the {\tt py-fmas} library code.  In
sects.~\ref{sec:soft_dep}-\ref{sec:soft_ext} we discuss the dependencies of the
software and software extendibility, and in sect.~\ref{sec:usage_examples} we
illustrate three usage examples.
Additional features of the provided software are discussed in the appendix.  In
particular, \ref{sec:spectrograms} discusses the capability to compute
spectrograms, \ref{sec:hR} details the various implemented Raman response
models, and \ref{sec:prop_const_class} illustrates a convenience class for
handling and analyzing propagation constants.
The {\tt py-fmas} code, a reference manual, an extended user guide and several
usage examples are available online at \url{\LINKTOCODE} \cite{pyfmas:gitHub}.

\section{Propagation models for the analytic signal \label{sec:prop_models}}

We here consider a periodic sequence of linearly polarized electromagnetic
pulses propagation along the $z$-direction of a one-dimensional dispersive
and nonlinear medium, supporting single-mode propagation
\cite{Amiranashvili:PRA:2010,Amiranashvili:AOT:2011}.  Let 
\begin{align}
E(z,t) = \mathsf{F}^{-1}\left[ E_\omega(z) \right] = \sum_\omega E_{\omega}(z) \,e^{-i\omega t},
                                 \quad \omega \in \frac{2 \pi}{T} \mathbb{Z},
                                                            \label{eq:t_rep_E}
\end{align}
denote the corresponding real-valued 
field in a temporal domain of period $T = 2 t_{\rm{max}}$, with
\begin{align}
E_\omega(z)  = \mathsf{F}\left[ E(z,t) \right] = \frac{1}{T} \int_{-t_{\rm{max}}}^{t_{\rm{max}}} 
                                    E(z,t) \, e^{i\omega t}~{\rm{d}}t, 
                                                            \label{eq:w_rep_E}
\end{align}
where  $E_\omega = E_{-\omega}^*$. 
The average field is considered to be constant along the $z$-direction,
assuming \mbox{$E_{\omega=0}(z)=0$}.
Above, $\mathsf{F}$ and $\mathsf{F}^{-1}$ denote the forward and inverse
Fourier transform.  
Then, under conditions that are common in optical fibers the field of the pulse
sequence can be described in terms of the nonlinear first-order
$z$-propagation equation \cite{Amiranashvili:PRA:2010,Amiranashvili:AOT:2011}
\begin{align}
i\partial_z E_\omega + [\beta(\omega)\!+\!i\alpha(\omega)] E_\omega +
                    \frac{\omega^2 \chi}{2 c^2 \beta(\omega)}\sum_{123|\omega}\! 
                                E_{\omega_1} E_{\omega_2}E_{\omega_3} = 0,
                                                                \label{eq:UA}
\end{align}
where $c$ is the speed of light, $\chi$ is a constant nonlinear susceptibility
specifying a cubic Kerr model, and the sum-index \mbox{token $123|\omega$} abbreviates
the condition $\omega_1+\omega_2+\omega_3 = \omega$.  
In Eq.~(\ref{eq:UA}), the frequency dependent propagation constant $\beta$ and
attenuation factor $\alpha$ specify the real-valued odd and even parts of the
complex valued wave number
$k(\omega)=\beta(\omega)+i\alpha(\omega)=\omega\sqrt{\epsilon(\omega)}/c$,
related to the dielectric constant $\epsilon$.  The propagation constant
relates to the refractive index
$n(\omega)=\mathsf{Re}[\sqrt{\epsilon(\omega)}]$ in the form
$\beta(\omega)=\omega n(\omega)/c$.  
%


Based on the above unidirectional propagation model for the field, a sequence
of simplified models can be derived that have the advantage of directly
neglecting non-resonant contributions of four-wave mixing in the nonlinear part
of Eq.~(\ref{eq:UA}).
This is achieved by considering, instead of the real optical field $E(z,t)$,
the complex-valued analytic signal 
\begin{align}
\Ecal(z,t)=\sum_{\omega>0} \Ecal_\omega(z)\,e^{i\omega t},\quad
    \Ecal_{\omega}(z)=\left[1+{\rm{sign}}(\omega)\right] E_\omega(z),
                                                            \label{eq:AS}
\end{align}
with $\Ecal_{\omega<0}=0$. 
The optical field is related to the analytic signal by
$E=(\Ecal\!+\!\Ecal^*)/2=\mathsf{Re}[\Ecal]$.  
%
Using Eq.~(\ref{eq:AS}), the propagation equation for the analytic signal,
derived from Eq.~(\ref{eq:UA}), reads \cite{Demircan:OE:2014}
\begin{align}
i\partial_z \Ecal_\omega + k(\omega) \Ecal_\omega +
                    \frac{\omega^2 \chi}{8 c^2 \beta(\omega)} 
                                \left((\Ecal+\Ecal^*)^3\right)_{\omega>0} = 0,
                                                           \label{eq:FMAS_full}
\end{align}
wherein $[\cdot]_{\omega>0}$ denotes spectral components restricted to the
positive frequency part of the nonlinearity.  In terms of this analytic signal,
all four-wave-mixing (FWM) processes that enter Eq.~(\ref{eq:UA}) can be
separated. Specifically we can separate the field product in the nonlinear part
according to
\begin{subequations}
\begin{align}
(\Ecal + \Ecal^*)^3~=~\quad
          &\Ecal^3 
                                                            \label{eq:FWM_1}\\
        +~ &3\,|\Ecal|^2\Ecal
                                                            \label{eq:FWM_2}\\
        +~ &3\,|\Ecal|^2\Ecal^*
                                                            \label{eq:FWM_3}\\
        +~ &\Ecal^{*3},        &
                                                            \label{eq:FWM_4}
\end{align}
\end{subequations}
where Eq.~(\ref{eq:FWM_1}) facilitates third-harmonic generation (THG),
Eq.~(\ref{eq:FWM_2}) is a Kerr-type nonlinear term,  Eq.~(\ref{eq:FWM_3}) is a
conjugate Kerr-type term,  and Eq.~(\ref{eq:FWM_4}) can be neglected when
restricting to $\omega>0$ in Eq.~(\ref{eq:FMAS_full}).
%
Considering only FWM processes defined by Eq.~(\ref{eq:FWM_2}),
Eq.~(\ref{eq:FMAS_full}) simplifies to
\cite{Amiranashvili:AOT:2011,Amiranashvili:PRA:2010}
\begin{align}
i\partial_z \Ecal_\omega + k(\omega) \Ecal_\omega +
                    \frac{3 \omega^2 \chi}{8 c^2 \beta(\omega)} 
                                \left(|\Ecal|^2 \Ecal\right)_{\omega>0} = 0,
                                                              \label{eq:FMAS}
\end{align}
which we here refer to as the forward model for the analytic signal (implemented
as model {\tt FMAS}).
Further, we here refer to Eq.~(\ref{eq:FMAS_full}) as the forward model for the
analytic signal including terms such as third-harmonic generation 
(implemented as model {\tt FMAS\_THG}).

%
Additional simplification of the nonlinear term is possible by approximating 
$\beta(\omega) \approx \omega  n(\omega_0)/c$ for a reference frequency $\omega_0$, 
and by expressing the nonlinear susceptibility $\chi$ through the nonlinear
refractive index $n_2$ as  $\chi = \frac{8}{3}n(\omega_0) n_2$. 
Then, Eq.~(\ref{eq:FMAS}) can be written in the form \cite{Demircan:OE:2014,Demircan:APB:2014}
\begin{align}
i\partial_z \Ecal_\omega + k(\omega) \Ecal_\omega +
                    n_2 \frac{\omega}{c} 
                                \left(|\Ecal|^2 \Ecal\right)_{\omega>0} = 0,
                                                              \label{eq:FMAS_S}
\end{align}
yielding a simplified forward model for the analytic signal
(implemented as model {\tt FMAS\_S}).

%
The Raman effect is incorporated by augmenting the nonlinear part in the
form \cite{Demircan:OE:2014,Demircan:APB:2014}
\begin{align}
i\partial_z \Ecal_\omega + k(\omega) \Ecal_\omega +
    n_2 \frac{\omega}{c}\left((1-f_R)\,|\Ecal|^2 \Ecal + f_R\,\Ecal \mathcal{I}_R \right)_{\omega>0} = 0,
                                                              \label{eq:FMAS_S_R}
\end{align}
where $f_R$ specifies the fractional Raman contribution, and 
\begin{align}
\mathcal{I}_R = \sum_\omega h(\omega) 
                        \left(|\Ecal|^2 \right)_\omega e^{-i\omega t},~
h(\omega)=\frac{\tau_1^{-2}+\tau_2^{-2}}{\tau_1^{-2}-(\omega +i \tau_2^{-1})^2},
                                                            \label{eq:I_Raman}
\end{align}
represents convolution with a generic two parameter Raman response function
$h(\omega)$.  The latter implements an approximation by a
single-damped-harmonic oscillator with parameters
$\tau_{1,2}$. For example, for silica fibers
adequate values are $f_R=0.18$, $\tau_1=12.2\,\mathrm{fs}$, and
$\tau_2=32\,\mathrm{fs}$.  More specific expressions for the response function
$h(\omega)$ might of course be used, see \ref{sec:hR}.  Equation (\ref{eq:FMAS_S_R}) comprises
the simplified forward model for the analytic signal including the Raman effect
(implemented as model {\tt FMAS\_S\_R}).

Subsequently, so as to assess the accuracy of our numerical simulations in the
no-loss limit ($\alpha=0$), we consider the conserved quantities
\begin{align}
C_p(z) = 
    \begin{cases}
\sum_{\omega>0} \omega^{-2}\,\beta(\omega)\,|\Ecal_\omega(z)|^2, &~\text{for Eqs.~(\ref{eq:FMAS_full},\ref{eq:FMAS})},\\
\sum_{\omega>0} \omega^{-1}\,|\Ecal_\omega(z)|^2, &~\text{for Eqs.~(\ref{eq:FMAS_S},\ref{eq:FMAS_S_R})},\\
    \end{cases}
    \label{eq:Cp}
\end{align}
which are related to the classical analog of the photon number, see
Refs.~\cite{Amiranashvili:PRA:2010,Amiranashvili:BOOK:2016} for
Eqs.~(\ref{eq:FMAS_full},\ref{eq:FMAS}), and
Refs.~\cite{Blow:JQE:1989,Conforti:OE:2013,Zheltikov:PRA:2018} for models with
nonlinearity of the form of Eqs.~(\ref{eq:FMAS_S},\ref{eq:FMAS_S_R}).

\section{Computational problem solved by the software\label{sec:problem}}

The computational problem solved by the provided software is an initial value
problem, consisting of the propagation of a complex-valued field
$\mathcal{E}(z,t)$ along the propagation coordinate $z$ on a periodic
$t$-domain of extend $T=2 t_{\rm{max}}$, governed by a nonlinear partial
differential equation (PDE) of first order, i.e.\
\begin{subequations}
  \begin{align}
    &\partial_z \mathcal{E}(z,t) = L\mathcal{E}(z,t) + 
                                   N\left(\mathcal{E}(z,t)\right), 
                                   && z\geq0, |t| \leq t_{\rm{max}},   
                                   \label{eq:PDE}\\
    &\mathcal{E}(z,-t_{\rm{max}}) = \mathcal{E}(z,t_{\rm{max}}) 
                                   && z\geq 0,                
                                   \label{eq:BCs}\\
    &\mathcal{E}(z, t)|_{z=0} = \mathcal{E}_0(t) 
                                   && |t| \leq t_{\rm{max}}.           
                                   \label{eq:IC}
  \end{align}
\end{subequations}
In Eq.~(\ref{eq:PDE}), $L$ and $N$ are linear and nonlinear operators,
respectively.  Equation~(\ref{eq:BCs}) specifies the boundary conditions, and
Eq.~(\ref{eq:IC}) specifies the initial condition.
Taking the Fourier transform of Eq.~(\ref{eq:PDE})  we obtain the equation
\begin{align}
  \partial_z \mathcal{E}_\omega(z) = \hat{L}(\omega) \mathcal{E}_\omega(z) + 
                                  \hat{N}(z),
                                  \label{eq:generic_PDE_w}
\end{align}
where $\Ecal_\omega(z)=\mathsf{F}\left[\Ecal(z,t) \right]$, $\hat{L}(\omega) =
i k(\omega)$, and $\hat{N}(z) = \mathsf{F}\left[
N\left(\Ecal(z,t)\right)\right]$.
The models defined by \mbox{Eqs.~(\ref{eq:FMAS_full}, \ref{eq:FMAS},
\ref{eq:FMAS_S}, \ref{eq:FMAS_S_R})}, are conveniently expressed in the generic
form of Eq.~(\ref{eq:generic_PDE_w}), see Tab.~\ref{tab:TAB1}, which allows for
effective pseudospectral implementations. The $z$-propagation algorithms
implemented in {\tt py-fmas} are discussion in sect.~\ref{sec:algs} below.

\begin{table}[htb]
\caption{
List of implemented models. Propagation equations are specified using the
frequency domain representation of their linear ($\hat{L}(\omega)$) and
nonlinear ($\hat{N}$) operators in the form of Eq.~(\ref{eq:generic_PDE_w}).
A conserved quantity, related to the classical expression for
the photon number, valid in the no-loss limit ($\alpha=0$), is implemented
by default ($C_p$).
}
\label{tab:TAB1}
\renewcommand{\arraystretch}{1.2} 
\begin{footnotesize}
\begin{tabular}{llll}
\hline
\hline
Model  &  $\hat{L}(\omega)$      & $\hat{N}(z)$ & $C_p(z)$\\
\hline
{\tt FMAS\_THG}  & $ik(\omega)$    
                 & $i\frac{3\omega^2\chi}{8 c^2 \beta(\omega)}\left((\Ecal+\Ecal^*)^3\right)_{\omega>0}$ 
                 & $\sum\limits_{\omega>0} \frac{\beta(\omega)}{\omega^2} |\Ecal_\omega|^2$
                 \\[0.2cm]
{\tt FMAS}       & $ik(\omega)$     
                 & $i\frac{3\omega^2\chi}{8 c^2 \beta(\omega)}\left(|\Ecal|^2 \Ecal\right)_{\omega>0}$ 
                 & $\sum\limits_{\omega>0} \frac{\beta(\omega)}{\omega^2} |\Ecal_\omega|^2 $
                 \\[0.2cm]
{\tt FMAS\_S}    & $ik(\omega)$     
                 & $i\frac{n_2\omega}{c}\left(|\Ecal|^2 \Ecal\right)_{\omega>0}$ 
                 & $\sum\limits_{\omega>0} \omega^{-1} |\Ecal_\omega|^2$
                 \\[0.2cm]
{\tt FMAS\_S\_R} & $ik(\omega)$     
                 & $i\frac{n_2 \omega}{c}\left((1\!-\!f_R)\,|\Ecal|^2 \Ecal\!+\!f_R\,\Ecal \mathcal{I}_R \right)_{\omega>0}$ 
                 & $\sum\limits_{\omega>0}  \omega^{-1}|\Ecal_\omega|^2$
                 \\[0.2cm]
                 &   
                 & $ \mathcal{I}_R\!=\!\sum\limits_{\omega}^{} h(\omega) \left(|\Ecal|^2 \right)_\omega e^{-i\omega t} $ 
                 &
                 \\[0.2cm]
\hline
\hline
\end{tabular}
\end{footnotesize}
\end{table}

\section{Implemented algorithms\label{sec:algs}}

A commonality of the above models is that their linear subproblem can be solved
by direct integration. 
That is, if only the linear part of Eq.~(\ref{eq:generic_PDE_w}) is nonzero, an
exact solution is given by $\mathcal{E}_\omega(z)= {\bf{P}}_{\rm{lin}}(z)\,
\mathcal{E}_\omega(0)$. In the latter,
\begin{align}
 {\bf{P}}_{\rm{lin}}(z) = e^{\hat{L}(\omega)\,z}\label{eq:P_lin}
\end{align}
is the exact linear propagator for advancing $\mathcal{E}_\omega$ under the
action of the linear operator. In this case, a solution to Eq.~(\ref{eq:PDE})
can be computed as $\mathcal{E}(z,t) = \mathsf{F}^{-1}\left[
{\bf{P}}_{\rm{lin}}(z)\, \mathsf{F}\left[ \mathcal{E}(0,t)\right]\right]$.
All propagation schemes implemented in {\tt py-fmas} module {\tt solver}
exploit the above property.

To advance a field for a single step along a discrete $z$-grid, three
fixed-stepsize algorithms are implemented. These are the simple split-step
Fourier method (SiSSM; sect.~\ref{sec:SiSSM}),
symmetric split-step
Fourier method (SySSM; sect.~\ref{sec:SySSM}), and
integrating factor method (IFM; sect.~\ref{sec:IFM}).  
{\tt py-fmas} also implements two adaptive stepsize algorithms, referred to as
the local error method (LEM; sect.~\ref{sec:LEM}) and the conservation quantity
error (CQE; sect.~\ref{sec:CQE}) method, where a single step of extend $\Delta
z$ possibly requires several substeps of the solver.  
Both methods aim at keeping the local error smaller than a prescribed error
bound by decreasing the stepsize when necessary while increasing the stepsize
when possible.
%

\subsection{Available $z$-stepping formulas\label{sec:z_stepper}}

A $z$-stepping formula implements the algorithmic procedure to advance a field
for a single step from position $z$ to $z+\Delta z$. 
This is important for solving the nonlinear subproblem of the considered models.
Let $z_n$ and $y_n$ be the $z$-position and field after step $n$, then taking a
single step can be abbreviated as 
\begin{align}
  y_{n+1} = {\bf{S}}(f, z_n, y_n, \Delta z),\label{eq:z_stepper} 
\end{align}
where $f = dy/dz$ is the evolution rate of the system to be solved and
$\Delta z$ is the step-size to be used.
{\tt py-fmas} provides functions implementing a second-order Runge-Kutta
formula (RK2; local error $\mathcal{O}(\Delta z^3)$) and fourth-order
Runge-Kutta formula (RK4; local error $\mathcal{O}(\Delta z^5)$)
\cite{NR:BOOK:2007}. 
%
In Eq.~(\ref{eq:z_stepper}), $y$ not necessarily refers to the
analytic signal. For example, the integrating factor method
(sect.~\ref{sec:IFM}) advances an auxiliary field that is different from the
analytic signal.

\subsection{Simple split-step Fourier method (SiSSM)\label{sec:SiSSM}}

In terms of the simple split-step Fourier method
\cite{Taha:JCP:1984,Weideman:SIAM:1986}, we advance a solution from $z$ to
$z+\Delta z$ by a subsequent composition of a nonlinear and a linear substep in
the form
\begin{subequations}
  \begin{align}
    \xi &= {\bf{S}}( \hat{N}, z, \mathcal{E}_\omega(z), \Delta z),
                                                        \label{eq:SiSSM_1}\\
    \mathcal{E}_\omega(z+\Delta z) &= {\bf{P}}_{\rm{lin}}(\Delta z)\,\xi.
                                                        \label{eq:SiSSM_2}
  \end{align}
\end{subequations}
The maximally achievable local error of this integration scheme is $O(\Delta
z^2)$. The maximally achievable global error, accumulated over the full
propagation range, is thus $O(\Delta z)$.  A solver based on the above method
is implemented as {\tt SiSSM} (in the text referred to as SiSSM). By default it
employs the RK2 $z$-stepping formula.

\subsection{Symmetric split-step Fourier method (SySSM)\label{sec:SySSM}}

For the symmetric split-step Fourier method \cite{DeVries:AIP:1987}, we advance
the solution from position $z$ to $z+\Delta z$ by a subsequent composition of a
linear half-step, a full nonlinear step, and a final linear half step in the
form
\begin{subequations}
  \begin{align}
    \xi &= {\bf{P}}_{\rm{lin}}(\Delta z/2)\,\mathcal{E}_\omega(z),
                                                      \label{eq:SySSM_1}\\
    \xi^{\prime} &= {\bf{S}}(\hat{N}, z,\xi, \Delta z),
                                                      \label{eq:SySSM_2}\\
    \mathcal{E}_\omega(z+\Delta z) &= {\bf{P}}_{\rm{lin}}(\Delta z/2)\,
                                            \xi^{\prime}. 
                                                      \label{eq:SySSM_3}
  \end{align}
\end{subequations}
This integration scheme yields a maximal achievable global error $O(\Delta
z^2)$.  A solver based on the above method is implemented as {\tt SySSM} (in
the text referred to as SySSM). By default it employs the RK2 $z$-stepping
formula.

\subsection{Integrating factor method (IFM)\label{sec:IFM}}

Starting with the generic partial differential equation in the frequency domain,
Eq.~(\ref{eq:generic_PDE_w}), we define the auxiliary fields
\begin{align}
  \phi_\omega(z) = {\bf{P}}_{\rm{lin}}(z_0-z)\,\mathcal{E}_\omega(z),
                                                         \label{eq:aux_fields}
\end{align}
where ${\bf{P}}_{\rm{lin}}(z_0-z)=\exp\{\hat{L}(\omega)\,(z_0-z)\}$ specifies
the integrating factor and $z_0$ is a reference position
\cite{Milewski:SIAM:1999,Kassam:SIAM:2005}.  Replacing $\mathcal{E}_\omega$ in
Eq.~(\ref{eq:generic_PDE_w}) by these auxiliary fields eliminates the linear
part and yields a system of ordinary differential equations
\cite{Milewski:SIAM:1999,Trefethen:BOOK:2000,Kassam:SIAM:2005}
\begin{subequations}
\begin{align}
   \partial_z \phi_\omega(z) &= {\bf{P}}_{\rm{lin}}(z_0-z)\,
                                \hat{N}\left(  
                                          {\bf{P}}_{\rm{lin}}(z-z_0)\,
                                          \mathcal{\phi}_\omega(z) 
                                        \right)\\
                             &=\hat{G}_{z_0}(z,\phi_\omega(z)),\label{eq:IP}
  \end{align}
\end{subequations}
coupled through the nonlinear function $\hat{G}_{z_0}$.  
%
%
Equation~(\ref{eq:IP}) defines an ``interaction picture'' representation of
Eq.~(\ref{eq:generic_PDE_w}): in absence of an ``interaction'', i.e.\ if only
the linear part of Eq.~(\ref{eq:generic_PDE_w}) is nonzero, the auxiliary
fields satisfy $\partial_z \phi_\omega=0$.
In order to advance the original field from $z$ to $z+\Delta z$, we here choose
the reference position $z_0=z+\Delta z/2$ in
Eqs.~(\ref{eq:aux_fields},\ref{eq:IP}). For this choice, the midpoint
derivative for the auxiliary field matches that of the original field.  A full
step of the integrating factor method is then given by the composition
\begin{subequations}
  \begin{align}
    \phi_\omega(z) &= {\bf{P}}_{\rm{lin}}(\Delta z/2)\,\mathcal{E}_\omega(z),
                                                            \label{eq:IFM_1}\\
    \xi &= {\bf{S}}( \hat{G}_{z_0=z+\Delta z/2}, z, \phi_\omega(z), \Delta z),
                                                            \label{eq:IFM_2}\\
    \mathcal{E}_\omega(z+\Delta z) &= {\bf{P}}_{\rm{lin}}(\Delta z/2)\,\xi.
                                                            \label{eq:IFM_3}
  \end{align}
\end{subequations}
Equation~(\ref{eq:IFM_1}) performs the change to the auxiliary field
at position $z$ using Eq.~(\ref{eq:aux_fields}), Eq.~(\ref{eq:IFM_2}) advances
the auxiliary field using Eq.~(\ref{eq:IP}), and Eq.~(\ref{eq:IFM_3}) recovers
the original field at $z+\Delta z$, again using Eq.~(\ref{eq:aux_fields}).  
%
For the above choice of the reference position $z_0$, the sequence of substeps
Eqs.~(\ref{eq:IFM_1}-\ref{eq:IFM_3}) has a structure similar to the SySSM
scheme [Eqs.~(\ref{eq:SySSM_1}-\ref{eq:SySSM_3})].
However, using the RK4 algorithm for $z$-stepping in Eq.~(\ref{eq:IFM_2})
results in a global error $O(\Delta z^4)$.
The above variant of the integrating factor method is referred to as the
``Runge-Kutta in the interaction picture'' (RK4IP) method \cite{Hult:JLT:2007}.
Such schemes are also referred to as linearly exact Runge-Kutta methods \cite{Archilla:JCP:1995}.
A solver based on the above procedure is implemented as {\tt IFM\_RK4IP} (in
the text referred to as IFM-RK4IP).

\subsection{Local-error method (LEM)\label{sec:LEM}}

The second-order accurate symmetric split-step method (SySSM;
sect.~\ref{sec:SySSM}) can be used to devise a refined algorithm with local
error $\mathcal{O}(\Delta z^4)$ \cite{Sinkin:JLT:2003}.  This can be
achieved by step-doubling and local extrapolation \cite{NR:BOOK:2007}.  To
advance a solution from position $z$ to $z+h$, step-doubling proceeds by
computing a coarse solution $\mathcal{E}_{\omega}^{\rm{(c)}}(z+h)$ using a full
step of extend $h$, and a fine solution $\mathcal{E}_{\omega}^{\rm{(f)}}(z+h)$
using a subsequent composition of two half-steps of extend $h/2$.
Local extrapolation consists of combining the results in the form
\cite{Sinkin:JLT:2003} 
\begin{align}
\mathcal{E}_\omega(z+h) = 
                    \frac{4}{3} \mathcal{E}_{\omega}^{\rm{(f)}}(z+h)  
                   -\frac{1}{3} \mathcal{E}_{\omega}^{\rm{(c)}}(z+h).
                                                        \label{eq:LEM_extrap}
\end{align}
In comparison to the number of evaluations of Eq.~(\ref{eq:SySSM_2}) needed to
compute the fine solution, the overhead cost for evaluating
Eq.~(\ref{eq:LEM_extrap}) is a factor $1.5$. 

In the local-error method \cite{Sinkin:JLT:2003}, the relative local error
\begin{align}
\delta_{\rm{RLE}} = \frac{  || \mathcal{E}_\omega^{\rm{(f)}} 
                            -\mathcal{E}_\omega^{\rm{(c)}} ||
                         }{ || \mathcal{E}_\omega^{\rm{(f)}} ||},
                                                        \label{eq:LEM_lre}
\end{align}
with norm $||x|| = \sqrt{\int |x|^2~{\rm{d}\omega}}$, is used to assess the
performance of the algorithm and to adapt the stepsize so that
$\delta_{\rm{RLE}}$ is kept within a target range $(\delta_{G}/2, \delta_G)$,
specified by a goal local error $\delta_G$ provided by the user.
Let us note that in order to advance a solution by one $z$-slice of extend
$\Delta z$, the local-error method performs possibly multiple substeps of
extend $h\leq \Delta z$.  The protocol for adapting the local step
size $h$ distinguishes three cases \cite{Sinkin:JLT:2003}:
\begin{enumerate}
\item If $\delta_{\rm{RLE}} > 2\,\delta_G$, discard the current trial solution
\mbox{$\mathcal{E}_\omega(z+h)$} and retry the substep with $h \leftarrow h/2$.
\item If $\delta_G < \delta_{\rm{RLE}} \leq 2\,\delta_G$, keep the trial
solution and decrease the stepsize to $h\leftarrow 2^{-1/3}\,h$ for the next
substep.
\item If $\delta_{\rm{RLE}} < \delta_G/2$, keep the trial solution and increase
the stepsize to $h \leftarrow 2^{1/3}\,h$ for the next substep.
\end{enumerate}
Otherwise, if the relative local error is within the target range specified 
by the goal local error, the trial solution and local stepsize are kept. 
The substep completing each $z$-slice is truncated to terminate exactly at
$z+\Delta z$. Thus, on termination of the algorithm, the field solution
$\mathcal{E}_\omega$ is available on a discrete $z$-grid with constant spacing
$\Delta z$.  
By default, the LEM algorithm is used in conjunction with a RK2
$z$-stepping formula. 
A solver based on the above procedure is implemented as {\tt LEM} (in the text
referred to as LEM).

\subsection{Conservation quantity error method (CQE) \label{sec:CQE}}

We here also provide an implementation of the conservation quantity error (CQE)
method \cite{Heidt:JLT:2009}, wherein stepsize adaption is controlled by a
conservation law of the underlying model equation.
To advance a solution from position $z$ to $z+h$, the integrating factor method
IFM-RK4IP (sect.~\ref{sec:IFM})
is used. By default, Eq.~(\ref{eq:Cp}) is used to guide stepsize adaption.
For this purpose, the relative photon number error \cite{Heidt:JLT:2009}
\begin{align}
\delta_{\rm{Ph}}(z)=\frac{|C_{{p}}(z+h)-C_{{p}}(z)|}{C_{{p}}(z)}
                \label{eq:CQE_err}
\end{align}
is monitored and compared to a user provided goal local 
error $\delta_G$. 
In terms of the CQE, 
the protocol for adapting the stepsize $h$ reads \cite{Heidt:JLT:2009}:
\begin{enumerate}
\item If $\delta_{\rm{Ph}} > 2\,\delta_G$, discard the current trial solution
\mbox{$\mathcal{E}_\omega(z+h)$} and retry the substep with $h \leftarrow h/2$.
\item If $\delta_G < \delta_{\rm{Ph}} \leq 2\,\delta_G$, keep the trial
solution and decrease the stepsize to $h\leftarrow2^{-1/5}\,h$ for the next
substep.
\item If $\delta_{\rm{Ph}} < 0.1\,\delta_G$, keep the trial solution and increase
the stepsize to $h\leftarrow2^{1/5}\,h$ for the next substep.
\end{enumerate}
Otherwise, if the relative local error is within the target range
$(0.1\,\delta_{G}, \delta_G)$, specified by the goal local error, the trial
solution and local stepsize are kept. 
A solver based on the above procedure, valid in the no-loss limit ($\alpha=0$),
is implemented as {\tt CQE}.  The conservation law used to control stepsize
adaption can be changed by the user by providing a suitable function when an
instance of the solver is initialized.  An example is provided
\href{\LINKTOCODE}{online} in the extended user guide \cite{pyfmas:gitHub}.

\begin{figure}[t!]
\includegraphics[width=\linewidth]{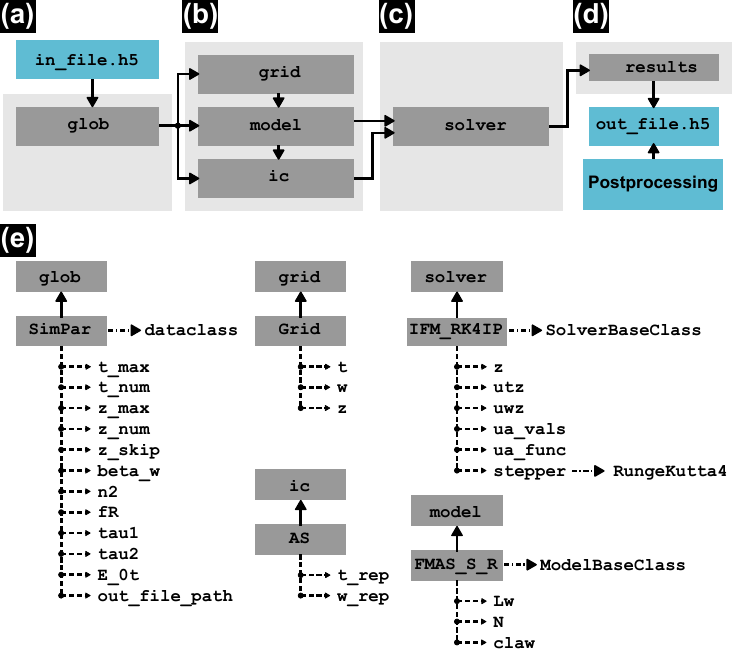}
\caption{
(a-d)  Pictorial outline of a workflow using the {\tt py-fmas} library code.
(a) Specification of a full simulation run.
(b) Initialization of the model.
(c) Initialization of the propagation scheme and $z$-propagation.
(d) Data storage.
Input data, output data and postprocessing tools are part of more ``specific
project'' code.
(e) Relationships between selected objects relevant to the minimal example
in listing~\ref{code:minimal_example}. 
In (e), solid arrows indicate class instantiation (``creates'' relationship),
dashed arrows indicate attributes and methods (``has-a'' relationship),
dash-dotted arrow indicates reference to an object (``is-a'' relationship).
\label{fig:FIG00}}
\end{figure}

\section{User manual \label{sec:user_manual}}

Below we clarify the structure of the {\tt py-fmas} python package and detail a
typical workflow involving the {\tt py-fmas} package.  Usage examples that
employ  the {\tt py-fmas} library code are discussed in
sect.~\ref{sec:usage_examples}.

\subsection{Structure of the {\tt py-fmas} package}

{\tt py-fmas} is a pure Python package, organized as a collection of modules.
The scope and capabilities of the individual modules are summarized below.
Further information is available in the
\href{https://omelchert.github.io/py-fmas/reference_manual/index.html}{online
reference manual} \cite{pyfmas:gitHub}.

\begin{itemize}
\item {\tt models}: Subpackage implementing the propagation models for the
analytic signal discussed in sect.~\ref{sec:prop_models} (see
Tab.~\ref{tab:TAB1}). Also provides a data structure allowing to implement
custom models.

\item {\tt solver}: Subpackage implementing the $z$-propagation algorithms
detailed in sect.~\ref{sec:algs}. Also provides a data structure allowing to
implement custom algorithms.

\item {\tt propagation\_constant}: Provides several propagation constants.
Also provides a data structure for analyzing user-defined propagation constants
(\ref{sec:prop_const_class}). 

\item {\tt raman\_response}: Provides functions that implement several 
Raman response models (\ref{sec:hR}).

\item {\tt stepper}: Provides the $z$-stepping formulas used by the propagation
algorithms (sect.~\ref{sec:z_stepper}).

\item {\tt analytic\_signal}: Provides data structures for converting
discrete-time optical field to discrete time analytic signal.

\item {\tt data\_io}: Provides functions and data structures for reading and
writing data in HDF5-format.

\item {\tt tools}: Provides functions for postprocessing (see
\ref{sec:spectrograms}) and visualizing simulation data, as well as functions
that did not fit into the other modules.

\item {\tt config}: Module containing functions and parameters jointly used
by several modules.

\end{itemize}

\subsection{Availability of the software}

{\tt py-fmas} is openly available \cite{pyfmas:gitHub}, hosted on the code
development platform gitHub (\url{https://github.com}). It is
implemented in Python3 under the MIT license. The software can be installed by
cloning the repository and installing the provided Python3 wheel:
\begin{footnotesize}
\begin{verbatim}
> git clone https://github.com/omelchert/py-fmas.git
> cd ./py-fmas/dist
> python3 -m pip install ./py_fmas-1.0-py3-none-any.whl
\end{verbatim}
\end{footnotesize}

\subsection{Specifying a propagation scenario \label{sec:workflow_a}}

While our computational research projects are usually carried out by scripting,
incorporating {\tt py-fmas} library code into more specific project code as
needed, we also provide convenience methods that allow a user to read a
propagation scenario from an input file in HDF5-format
[Fig.~\ref{fig:FIG00}(a)].
This file must contain all necessary simulation parameters for specifying the
computational domain, propagation model, and, propagation algorithm. All
required parameters are listed in Tab.~\ref{tab:sim_pars}. 
Let {\tt in\_file.h5} be an adequate input file, then a workflow can be started
by importing the {\tt fmas} package (listing~\ref{code:minimal_example}, line
1), and reading the simulation parameters using the function {\tt read\_hd5()}
contained in module {\tt data\_io} (listing~\ref{code:minimal_example}, line
3).
The obtained data structure {\tt glob} is an instance of the dataclass {\tt
SimPars}, implemented in {\tt data\_io} as well. 
The various parameters held by {\tt SimPars} [Fig.~\ref{fig:FIG00}(e)] are 
detailed in Tab.~\ref{tab:sim_pars} in \ref{sec:input}.
This completes the first stage of the workflow outlined in
Fig.~\ref{fig:FIG00}(a).
Let us note that the minimal example in listing~\ref{code:minimal_example} does
not use all of all the parameters implemented by the dataclass. A more general
application-type example that does this is discuss in sect.~\ref{sec:app}
below.
An example, demonstrating how to generate an adequate input file, is provided
along with the \href{\LINKTOCODE}{online documentation} \cite{pyfmas:gitHub}.

\subsection{Initializing a model}

Once a propagation scenario is specified, the problem specific data structures
can be initialized [Fig.~\ref{fig:FIG00}(b)].  First, a computational grid,
called {\tt grid}, is obtained as instance of the class {\tt Grid}
(listing~\ref{code:minimal_example}, lines 5--9).  {\tt Grid} provides
attributes for convenient access to the discrete coordinate axes
[Fig.~\ref{fig:FIG00}(e)].
For example, {\tt grid.t} returns a {\tt numpy.ndarray} of length
$M=t_{\rm{num}}$, defining the temporal grid points
\begin{subequations}
\begin{align}
t_m = -t_{\rm{max}} + m\,\Delta t,\quad m = 0,\ldots,M-1,
                                                            \label{eq:grid_t}
\end{align}
with $\Delta t = 2 t_{\rm{max}}/t_{\rm{num}}$ available as {\tt grid.dt}.
Likewise, {\tt grid.w} returns an array containing the angular frequency grid
points in standard order \cite{Scipy,NR:BOOK:2007}, i.e.
\begin{align}
\omega_m = 
  \begin{cases}
     m\, \Delta \omega, & ~\text{for}~m=0,\ldots,\frac{M}{2}-1,\\
   (m-M)\, \Delta \omega, & ~\text{for}~m=\frac{M}{2},\ldots,M-1,
  \end{cases}
                                              \label{eq:grid_w}
\end{align}
with $\Delta \omega = \pi/t_{\rm{max}}$ available as {\tt grid.dw}, 
and {\tt grid.z} returns an
array of length $z_{\rm{num}}+1$, containing the grid points
\begin{align}
z_n = 0 + n\,\Delta z, \quad n=0,\ldots,z_{\rm{num}},
                                                            \label{eq:grid_z}
\end{align}
\end{subequations}
along the propagation axis $z$, where the extend of a single $z$-slice $\Delta
z = z_{\rm{max}}/z_{\rm{num}}$ is available as {\tt grid.dz}.

Next, one of the propagation models implemented in module {\tt models}, namely
the simplified forward model including the Raman effect ({\tt FMAS\_S\_R}), is
initialized [listing~\ref{code:minimal_example}, lines 11--17;
Fig.~\ref{fig:FIG00}(e)], and the initial condition $E_0(t_m)$,
$m=0,\ldots,M-1$, is used to initialize a data structure holding the
corresponding discrete-time analytic signal
[listing~\ref{code:minimal_example}, line 19; Fig.~\ref{fig:FIG00}(e)].

\begin{lstlisting}[
    floatplacement = H, 
    numbers = left,
    captionpos = t,
    keywordstyle = \bf,
    frame = lines,
    stepnumber = 1,
    numbers = left, 
    numbersep = 5pt, 
    xleftmargin = \parindent,
    language = Python,
    basicstyle = \ttfamily\scriptsize, 
    caption = {Exemplary workflow using the {\tt py-fmas} library code.}, 
    label = code:minimal_example]
import fmas

glob = fmas.data_io.read_h5('in_file.h5')

grid = fmas.grid.Grid(
    t_max = glob.t_max,
    t_num = glob.t_num,
    z_max = glob.z_max,
    z_num = glob.z_num)

model = fmas.models.FMAS_S_R(
    w = grid.w,
    beta_w = glob.beta_w,
    n2 = glob.n2,
    fR = glob.fR,
    tau1 = glob.tau1,
    tau2 = glob.tau2)

ic = fmas.analytic_signal.AS(glob.E_0t)

solver = fmas.solver.IFM_RK4IP(
    model.Lw, model.Nw,
    user_action = model.claw)
solver.set_initial_condition(
    grid.w, ic.w_rep)
solver.propagate(
    z_range = glob.z_max,
    n_steps = glob.z_num,
    n_skip = glob.z_skip)

res = {
    "t": grid.t,
    "z": solver.z,
    "w": solver.w,
    "AS_tz": solver.utz,
    "Cp": solver.ua_vals}
fmas.data_io.save_h5('out_file.h5', **res)
\end{lstlisting}

\subsection{Initializing a solver and running a simulation}
Now that computational grid, model, and initial condition are set up, a
specific $z$-propagation algorithm can be initialized
[Fig.~\ref{fig:FIG00}(c)]. 
In lines 21ff, the minimal example in listing~\ref{code:minimal_example} shows
how to initialize an instance of the IFM-RK4IP solver
[Fig.~\ref{fig:FIG00}(e)], implemented in module {\tt solver}. In line 22, the
frequency-domain representation of the linear and nonlinear operators are
handed over. In line 23, an additional user-specified callback function is
initialized, that will be evaluated at each $z$-step.
Internally it is assigned to the class method {\tt ua\_fun}.
If a callback function is provided it needs to exhibit an interface
of the form \mbox{{\tt my\_fun(idx, zcurr, w, uw)}}, where {\tt idx} (type {\tt
int}) labels the current $z$-position {\tt zcurr} (type {\tt float}), {\tt w}
is the angular frequency grid (type {\tt numpy.ndarray}), and {\tt uw} is the
frequency domain representation of the field at the current $z$-position (type
{\tt numpy.ndarray}).
Here, {\tt model.claw} is handed over as callback function.  For the model
specified in line 11, the method {\tt claw} implements the conserved quantity 
Eq.~(\ref{eq:Cp}).

%
In line 24f, the initial condition is set. Let us note that all $z$-propagation
algorithms implemented in {\tt fmas} predominantly work in the frequency
domain. Hence, when setting the initial condition for the solver, a design
decision was to hand over the frequency-domain representation of the
discrete-time analytic signal.  In the second argument of the method call {\tt
set\_initial\_condition} this is achieved by {\tt ic.w\_rep}, which
implements the frequency-domain algorithm  \cite{Marple:TSP:1999}
\begin{align}
\mathcal{E}_{\omega_m} =
  \begin{cases}
    E_{\omega_m},    & m=0,\\
    2E_{\omega_m},   & 1\leq m \leq M/2-1,\\
    E_{\omega_{M/2}},  & m = M/2,\\
    0,              & M/2+1 \leq m \leq M - 1,
  \end{cases}
\end{align}
computing the frequency-domain representation of the discrete-time analytic
signal ($\mathcal{E}_{\omega_m}$, $m=0,\ldots,M-1$), based on the
frequency-domain representation of the real optical field ($E_{\omega_m}$,
$m=0.\ldots,M-1$).
A simulation run is best started with ``consistent'' initial conditions
that satisfy the boundary condition Eq.~(\ref{eq:BCs}). For example, for 
a propagation scenario starting off from a localized field pulse $\Ecal_0(t)$, the 
extend $T=2 t_{\rm{max}}$ of the periodic time domain should be set large
enough so that $|\Ecal_0(\pm t_{\rm{max}})|\approx 0$.

In lines 26--29 the algorithm is started, at which point the propagation range
({\tt z\_range}) and number of integration steps ({\tt z\_steps}) are
specified. The additional parameter {\tt n\_skip} specifies the number of
$z$-steps that are skipped in between two stored field configurations. This
allows the user to reduce the amount of output data generated by the solver.
While executing, the solver will evaluate the right-hand-side terms of 
Eq.~(\ref{eq:generic_PDE_w}) to perform the numerical integration along the 
\mbox{$z$-grid} and it will evaluate the optional user-specified callback-function,
if this is scheduled as shown in line 23.
The generated data will be stored on a discrete grid with grid
points
%
$z^\prime_n = n\,\Delta z^\prime$, with $\Delta z^\prime =
z_{\rm{max}}/N^\prime$, where $n = 0, \ldots, N^\prime$ and $N^\prime =
z_{\rm{num}}/z_{\rm{skip}}$.

Once the algorithm terminates, the time-domain representation of the
discrete-time analytic signal, given by $\mathcal{E}(z^\prime_n, t_m)$, with
$n=0,\ldots,N^\prime$, and $m=0,\ldots M-1$, is available as a two-dimensional
{\tt numpy.ndarray} retrieved by calling {\tt solver.utz}.
Likewise, $\mathcal{E}_{\omega_m}(z^\prime_n)$, $n=0,\ldots,N^\prime$,
$m=0,\ldots M-1$, can be retrieved by calling {\tt solver.uwz}.
The conserved quantity $C_p(z^\prime_n)$, with $n=0,\ldots,N^\prime$, is
available as one-dimensional {\tt numpy.ndarray} upon calling {\tt
solver.ua\_vals}.
Finally, the reduced $z$-grid with grid points $z^\prime_n$,
$n=0,\ldots,N^\prime$, is available as one-dimensional {\tt numpy.ndarray} upon
calling {\tt solver.z}.  Let us note that {\tt solver.z} and {\tt grid.z} have
the same length only for ${\tt n\_skip}=1$. 
Subsequently, we will refer to the coordinates $z^\prime$, at which a field
solution is given, simply as $z$.

\subsection{Data storage}

After the $z$-propagation algorithm has terminated, the generated data can be
saved to an output file in HDF5-format [Fig.~\ref{fig:FIG00}(d)].  
For this purpose, we set up a dictionary containing a {\tt key:data}-pair 
with custom key for each data object we want to save (lines 31--36), and
pass it to function {\tt save\_h5} provided by module {\tt data\_io} (line 37).

\subsection{Using {\tt fmas} as application \label{sec:app}}

The {\tt py-fmas} package can also be used as an application which interprets
all the attributes implemented by the dataclass {\tt SimPars}, see
Fig.~\ref{fig:FIG00}(e) and Tab.~\ref{tab:sim_pars}.  Such functionality is
implemented by the function {\tt run} in module {\tt app}.  
A minimal interactive python session that uses the {\tt fmas} library code
as an app reads:
\begin{verbatim}
>>> import fmas
>>> fmas.run('in_file.h5')
\end{verbatim}

A step-by-step example, demonstrating how to use {\tt py-fmas} as a black-box
application, choose a specific propagation model and algorithm, save data, and
generate a simple figure of the output is provided along with the
\href{\LINKTOCODE}{online documentation} \cite{pyfmas:gitHub}.

\subsection{Data postprocessing}
For subsequent analysis it is useful to consider the transformed field
$\mathcal{E}^\prime_\omega(z)=\mathcal{E}_\omega(z)\exp(i \omega z/v_0)$,
shifted to a moving frame of reference.  The time-domain representation
$\mathcal{E}^\prime(z,t)$ then corresponds to the analytic signal
$\mathcal{E}(z,\tau)$ for the retarded time $\tau =t-z/v_0$.  The reference
velocity $v_0$ can be chosen so that the time-domain dynamics appears slow.
For changing the frame of reference in this way, module {\tt tools} provides a function with interface 
\begin{verbatim}
change_reference_frame(w, z, uwz, v0)
\end{verbatim}
where {\tt w} (type {\tt numpy.ndarray}) is the $\omega$-grid, {\tt z} (type
{\tt numpy.ndarray}) is the $z^\prime$-grid, {\tt uwz} (type {\tt
numpy.ndarray}) is the two-dimensional freqnecy-domain representation of the
analytic signal
$\Ecal_\omega(z)$, and
{\tt v0} (type {\tt float}) is the reference velocity $v_0$.

Generally, generation of input data, postprocessing of output data and data
visualization is part of more specific project code and is not covered by the
minimal example discussed above. 
However, in module {\tt tools}, {\tt py-fmas} features simple functions that
assist a user to quickly visualize the generated data. 
%
As demonstrated in sect.~\ref{sec:use_case_3}, {\tt py-fmas} also includes the
functionality to compute simple spectrograms.
%

\section{Software dependencies \label{sec:soft_dep}}

{\tt py-fmas} is provided as a Python3 package \cite{Python}. It uses a wide
range of standard-library Python packages. The dependencies of {\tt py-fmas}
include:
\begin{itemize}
\item The Numpy and Scipy packages for python \cite{Scipy, Virtanen:NM:2020}.
\item The Matplotlib for data visualization \cite{Matplotlib}.
\item The HDF5 C-library for reading and writing files in HDF5 format, and
its Python wrapper h5py \cite{HDF5,h5py}.
\end{itemize}

\section{Software documentation \label{sec:soft_doc}}
{\tt py-fmas} is openly available. The
online documentation includes a reference manual with details on the
implemented models and propagation algorithms, an extended user guide with
step-by-step demonstrations of the functionality of {\tt py-fmas}, and further
usage examples. Links to the code-repository and the documentation are
available under \url{\LINKTOCODE} \cite{pyfmas:gitHub}.

\section{Software extendibility \label{sec:soft_ext}}

{\tt py-fmas} is based on our research code and was implemented with the aim of
beeing easily extendible and maintainable. 
%
For example, by using the base class {\tt ModelBaseClass}, contained in module
{\tt models}, it is straight-forward to implement further $z$-propagation models
for use with the {\tt py-fmas} library code. An example that shows how an
envelope model, given by the usual nonlinear Schr\"odinger equation (used in
sect.~\ref{sec:use_case_1}), can be set up is provided along with the
\href{\LINKTOPAGES}{online documentation} \cite{pyfmas:gitHub}.
There, we also demonstrate how to extend {\tt py-fmas} by models implementing
the Korteweg-deVries equation \cite{Zabusky:PRL:1965} and the Lugiato-Lefever
equation \cite{Lugiato:PRL:1987,Melchert:OL:2020}.  
In addition, we show how the provided software can be used to simulate
backscattered components of the optical field in terms of a bidirectional model
for a complex field \cite{Amiranashvili:PRA:2010}.  While this lies well within
the capabilities of the provided software, it is outside the
intended application domain of {\tt py-fmas}.
%
Further $z$-propagation schemes can be implemented via the {\tt
SolverBaseClass} provided in module {\tt solver}.
For example, implementing the ``Embedded Runge-Kutta scheme for step-size control
in the interaction picture method'' (ERK4(3)-IP)
\cite{Balac:CPC:2013,Balac:CPC:2016}, for use with analytic signal based
models,  is directly possible.
%
Another possibility is to extend the functionality of {\tt py-fmas} by the {\tt
optfrog} Python package \cite{Melchert:SFX:2019}, allowing to compute
analytic signal spectrograms with optimized time and frequency resolution.  An
example that illustrates this is available \href{\LINKTOCODE}{online}
\cite{pyfmas:gitHub}.

\section{Usage examples \label{sec:usage_examples}}

Below we show three use-cases of the software.
In sect.~\ref{sec:use_case_1} we demonstrate the accuracy of the implemented
algorithms when applied to the single soliton problem of the standard nonlinear
Schr\"odinger equation (NSE)
\cite{Miles:SIAM:1981,Taha:JCP:1984,DeVries:AIP:1987}, i.e.\ a test-case for
which an exact solution is available.
In sect.~\ref{sec:use_case_2} we reproduce an exemplary simulation of
supercontinuum generation in a photonic crystal fiber (PCF)
\cite{Demircan:OC:2005,Demircan:APB:2007,Dudley:RMP:2006,Hult:JLT:2007,Heidt:JLT:2009},
and compare the performance of fixed and adaptive stepsize schemes.
In sect.~\ref{sec:use_case_3} we show a complex propagation scenario involving
multiple interacting pulses at different center frequencies, out of the range
of the generalized nonlinear Schr\"odinger equation (GNSE).


\begin{figure}[t!]
\includegraphics[width=\linewidth]{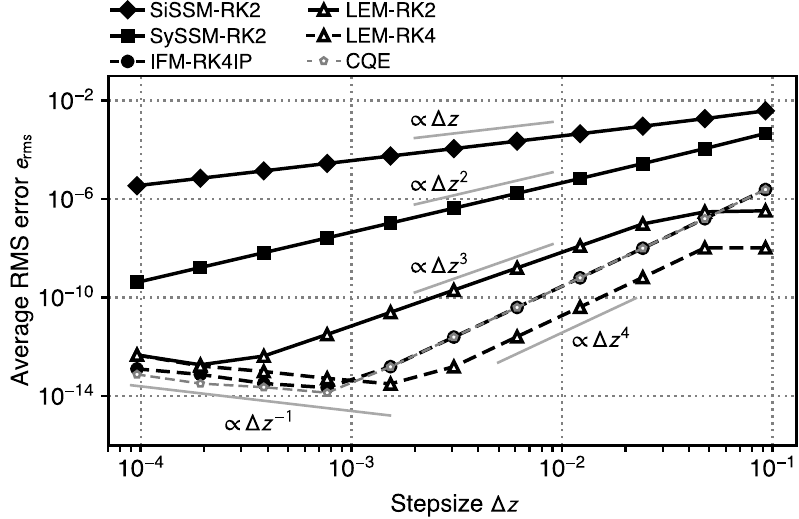}
\caption{
Average root-mean square (RMS) error of the different propagation schemes as
function of the stepsize $\Delta z$. Data is shown for different solver types
(SiSSM: Simple split-step Fourier method; SySSM: Symmetric split-step Fourier
method; IFM-RK4IP: Variant of integrating factor method aka.\ ``Runge-Kutta in
the interaction picture'' method; LEM: Local error method; CQE: Conservation
quantity error method) and $z$-stepping formulas (RK2: second-order Runge-Kutta
formula, solid lines; RK4: fourth-order Runge-Kutta formula, dashed lines). 
\label{fig:FIG01}}
\end{figure}

\subsection{Exact single-soliton solution of the NSE \label{sec:use_case_1}}
We first demonstrate that the functionality of the software can be extended 
by implementing additional models. 
Specifically, we here consider an envelope model given by the usual nonlinear
Schr\"odinger equation in the form \cite{Mitschke:BOOK:2010,Miles:SIAM:1981}
\begin{align}
\partial_z A = i \frac{1}{2} \partial_t^2 A + i |A|^2 A,\label{eq:NSE}
\end{align}
for the slowly varying complex pulse amplitude $A\equiv A(z,t)$, where, for
clarity, $z$ and $t$ are treated as dimensionless coordinates. 
The exact single-soliton solution of Eq.~(\ref{eq:NSE})
\cite{Miles:SIAM:1981,Drazin:BOOK:1989,Kivshar:BOOK:2003}, for unit pulse-width
given by
\begin{align}
A_{\rm{sol}}(z,t) = {\rm{sech}}(t)\, e^{-i z/2}, \label{eq:NSE_soliton} 
\end{align}
offers a possibility to assess the
accuracy of the implemented $z$-stepping algorithms.
Subsequently, we consider Eq.~(\ref{eq:NSE}) with initial condition
$A(0,t)={\rm{sech}}(t)$.  For the computational domain we choose
$t_{\rm{max}}=40$, $t_{\rm{num}}=4096$, and propagate for one soliton period,
i.e.\ up to $z_{\rm{max}}= \pi/2$, using different step sizes $\Delta z$.
Although the NSE allows for more specific implementations of split-step Fourier
methods that rely on an exact solution of the nonlinear subproblem, we here opt
to solve Eq.~(\ref{eq:NSE}) using the algorithms specified in
sect.~\ref{sec:algs}.
In Fig.~\ref{fig:FIG01} we show the resulting average root-mean square error (rms-error)
\begin{align}
e_{\rm{rms}} = \sqrt{\langle|A(z_{\rm{max}},t)-A_{\rm{sol}}(z_{\rm{max}},t)|^2\rangle} \label{eq:RMS_error}
\end{align}
at the final $z$-position as function of $\Delta z$. 
In Eq.~(\ref{eq:RMS_error}), $A(z,t)$ specifies the result of the numerical
integration at a given step size $\Delta z$. For all propagation schemes, a
scaling behavior \mbox{$e_{\rm{rms}}(\Delta z) = C \Delta z^{r}$}, for $\Delta z >
10^{-3}$ is clearly evident ($r$ denotes the scaling order of the rms-error,
see Fig.~\ref{fig:FIG01}).
We find the expected scaling of the global error for the different propagation
algorithms down to a saturation at \mbox{$e_{\rm{rms}} \approx 10^{-13}$}.
As evident from Fig.~\ref{fig:FIG01}, for this test problem, the local error
method (LEM) with a RK4 $z$-stepping formula exceeds the naively expected
$\mathcal{O}(\Delta z^3)$ behaviour by achieving an effective scaling 
$\mathcal{O}(\Delta z^4)$, a fortunate fact previously also noted in
Ref.~\cite{Heidt:JLT:2009}. Using a RK2 formula for $z$-stepping yields the
expected $\mathcal{O}(\Delta z^{3})$ error bound.
For very small values of $\Delta z$, the scaling behavior for the IFM and LEM
algorithms is $\propto \Delta z^{-1}$, i.e.\  proportional to the number of
performed Fourier-transforms.  
Since $\Delta z$ is small, this implies an overall large number of additions
and multiplications.  Consequently, the error scaling is dominated by the
accumulated round-off error. 
%
%

While the NSE provides a valuable testbed for assessing the performance of the
implemented algorithms, its ability to describe the dynamical evolution of
spectrally broad, ultrashort optical pulses is limited
\cite{Agrawal:BOOK:2013,Oughstun:PRL:1997,Zheltikov:PRA:2018}

\begin{table}[b]
\caption{
Expansion coefficients $\beta_n$ (in units of $\mathrm{fs^n/\mu m}$) for the polynomial
approximation of the propagation constant in Eq.~(\ref{eq:use_case_2_beta}).
Values are derived from the data shown in Tab.~1 of Ref.~\cite{Dudley:RMP:2006}.}
\label{tab:TAB2}
\renewcommand{\arraystretch}{1.2} 
\begin{center}
\begin{footnotesize}
\begin{tabular}{rS[table-format=1.6] l rS[table-format=1.5] l rS[table-format=1.4]}
\hline
\hline
$n$ & $\beta_n\,\mathrm{\left(\frac{fs^n}{\mu m}\right)}$ & &$n$ &   $\beta_n\,\mathrm{\left(\frac{fs^n}{\mu m}\right)}$  & &$n$ & $\beta_n\,\mathrm{\left(\frac{fs^n}{\mu m}\right)}$  \\
\cline{1-2} \cline{4-5} \cline{7-8}
2 &  -0.011830 &&  5 &   0.20737 && 8 &  -2.5495  \\ 
3 &   0.081038 &&  6 &  -0.53943 && 9 &   3.0524  \\
4 &  -0.095205 &&  7 &   1.34860 && 10 & -1.7140 \\
\hline
\hline
\end{tabular}
\end{footnotesize}
\end{center}
\end{table}

\begin{figure}[t!]
\includegraphics[width=\linewidth]{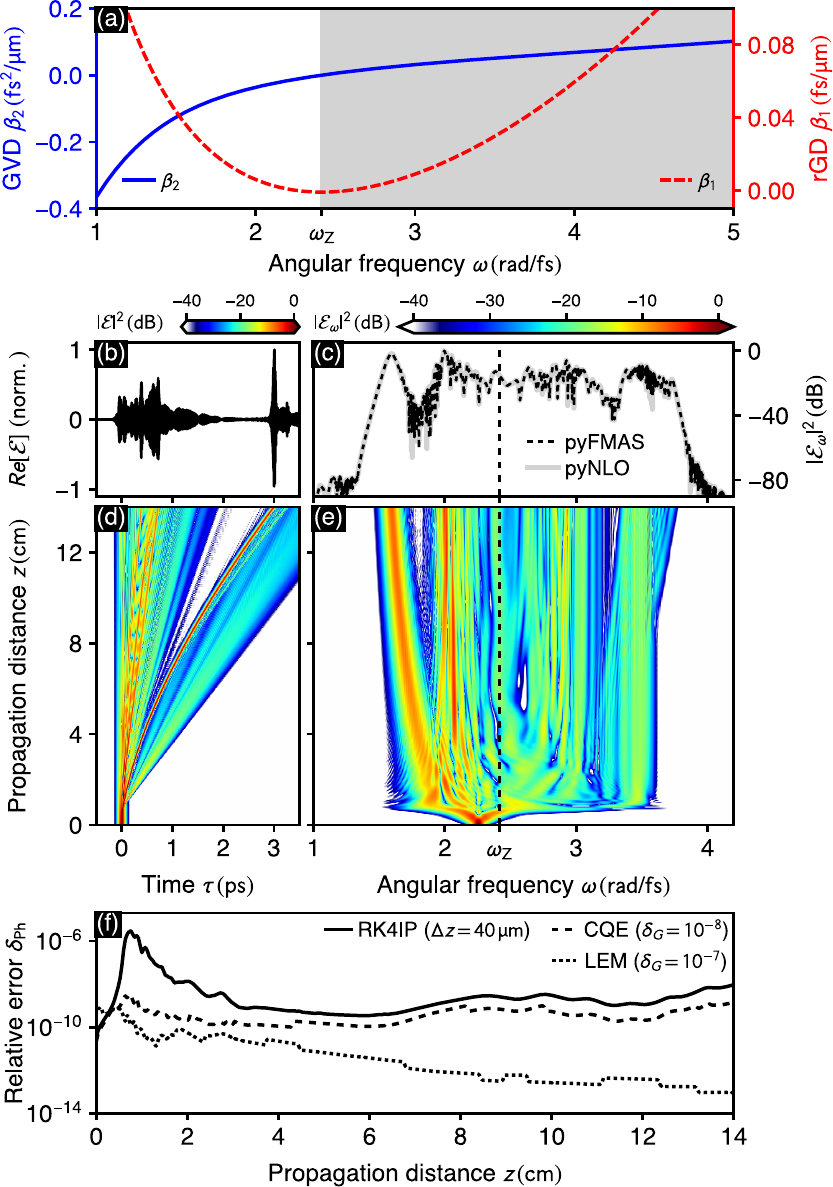}
\caption{
Exemplary simulation of supercontinuum generation in a PCF. 
(a) Frequency dependence of the relative group-delay (rGD) $\beta_1(\omega)$,
and group-velocity dispersion (GVD) $\beta_2(\omega)$.  Shaded region indicates
domain of normal dispersion with zero dispersion point $\omega_{\rm{Z}}\approx
2.415~\mathrm{rad/fs}$.
(b) Optical field $E=\mathsf{Re}[\mathcal{E}]$ at $z=14~\mathrm{cm}$. 
(c) Spectrum $|\mathcal{E}_\omega|^2$ at $z=14~\mathrm{cm}$.  
Solid line indicates results obtained using pyNLO \cite{pyNLO}. 
Dashed line shows results obtained using {\tt py-fmas} \mbox{IFM-RK4IP} solver
with stepsize $\Delta z=40\,\mathrm{\mu m}$. 
Evolution of (d) intensity, and, (e) spectrum over the length of the PCF.
Vertical dashed line in (c,e) indicates zero-dispersion point.
(f) Relative photon number error $\delta_{\rm{Ph}}$ obtained for the 
\mbox{IFM-RK4IP} solver, the local error method (LEM) and the conservation 
quantity error method (CQE).
\label{fig:FIG02}}
\end{figure}

\subsection{Supercontinuum generation in a PCF \label{sec:use_case_2}}

Next, we reproduce numerical results of a supercontinuum
generation process in a photonic crystal fiber (PCF).
The underlying propagation scenario is discussed on several occasions in the
scientific literature \cite{Dudley:RMP:2006,Hult:JLT:2007,Heidt:JLT:2009}.  For
example, in Ref.~\cite{Dudley:RMP:2006}, it is used to demonstrate numerical
simulations in terms of the generalized nonlinear Schr\"odinger equation
(GNSE) using the split-step Fourier method. 
In Ref.~\cite{Hult:JLT:2007} it is used to introduce the ``Runge-Kutta in
the interaction picture'' (RK4IP) method (sect.~\ref{sec:IFM}). 
In Ref.~\cite{Heidt:JLT:2009} it is used to demonstrate the feasibility of the
conservation quantity error method (CQE; sect.~\ref{sec:CQE}) for the
simulation of supercontinuum generation in optical fibers.
To investigate the sensitivity of the supercontinuum generation process on
different kinds of instabilities, an eighth-order Runge-Kutta scheme with
adaptive stepsize control has been used to ensure a high accuracy
\cite{Demircan:OC:2005,Demircan:APB:2007}.
%
All the above simulation studies used the GNSE, which relies on the slowly
varying envelope approximation. 
In contrast to this, we here employ also a
non-envelope model given by the simplified forward model for the analytic
signal with added Raman effect (FMAS-S-R).  
%
Specifically, we consider Eq.~(\ref{eq:FMAS_S_R}) with 
a polynomial approximation of the propagation constant, given by
\begin{align}
\beta(\omega) = \sum_{n=2}^{10} \frac{\beta_n}{n!} (\omega-\omega_0)^n, \label{eq:use_case_2_beta}
\end{align}
with parameters $\beta_n$ listed in Tab.~\ref{tab:TAB2},
$\omega_0=\,\mathrm{rad/fs}$ and $n_2=\gamma c/\omega_0$ with $\gamma= 0.11\!\cdot\!10^{-6}\,\mathrm{W^{-1}m^{-1}}$.  The frequency dependence of the relative
group delay $\beta_1(\omega)=\partial_\omega \beta(\omega)$ [note that using
Eq.~(\ref{eq:use_case_2_beta}) $\beta_1(\omega_0)=0\,\mathrm{fs/\mu m}$], and
group-velocity dispersion $\beta_2(\omega)=\partial_\omega^2 \beta(\omega)$ are
shown in Fig.~\ref{fig:FIG02}(a). The parameters specifying the Raman effect
are set to $f_R=0.18$, $\tau_1=12.2\,\mathrm{fs}$, and
$\tau_2=32.\,\mathrm{fs}$.
As initial condition we use
\begin{align}
E(0,t) = \mathsf{Re}\left[\sqrt{P_0}\,{\rm{sech}}\left(t/t_0 \right)\,e^{-i\omega_0 t} \right], \label{eq:use_case_2_iniCond}
\end{align}
with $P_0 = 10\,\mathrm{kW}$, and $t_0=28.4\,\mathrm{fs}$.  
%
For the computational domain we choose $t_{\rm{max}}=3.5\,\mathrm{ps}$,
$t_{\rm{num}}=2^{14}$, and propagate up to $z_{\rm{max}}= 14\,\mathrm{cm}$
using step size $\Delta z=40\,\mathrm{\mu m}$.
For $z$-propagation we use the IFM-RK4IP method.
The propagation dynamics of the field in both, time domain and frequency
domain, is detailed in Figs.~\ref{fig:FIG02}(b-e).  
In Fig.~\ref{fig:FIG02}(b,d), the pulse dynamics is shown as function of the
retarded time $\tau = t - \beta_1(\omega_0) z$.
In Fig.~\ref{fig:FIG02}(c)
we compare the analytic signal spectrum $|\Ecal_\omega|^2$ at
$z=14\,\mathrm{cm}$ to results obtained using the pyNLO code \cite{pyNLO}. 
Both agree well on a qualitative basis.
%
Figure \ref{fig:FIG02}(f) shows the relative photon number error defined in
Ref.~\cite{Heidt:JLT:2009}, related to the conserved quantity in
Eq.~(\ref{eq:Cp}) through $\delta_{\rm{Ph}}(z)=|C_{{p}}(z+\Delta
z)-C_{{p}}(z)|/C_{{p}}(z)$.
For the simulation run with stepsize $\Delta z = 40\,\mathrm{\mu m}$, the
maximum photon number error is $\delta_{\rm{Ph}}\approx 2.9\!\cdot\!10^{-6}$ at
$z\approx 0.76\,\mathrm{cm}$ [see Fig.~\ref{fig:FIG02}(f)]. We can compare this
to the results shown in Fig.~1(d) of Ref.~\cite{Heidt:JLT:2009}, exhibiting the
somewhat larger peak photon error of $\delta_{\rm{Ph}}\approx
6\!\cdot\!10^{-6}$.
In Fig.~\ref{fig:FIG02}(f) we also included the relative photon number error
obtained from a simulation run using the local error method (LEM), with local
goal error set to $\delta_G=10^{-7}$, and the conservation quantity error
method (CQE; $\delta_G=10^{-8}$). Here, the advantage of the adaptive stepsize
schemes is clearly evident. During the early propagation stage, i.e.\ for
$z<2\,\mathrm{cm}$, a decreased stepsize prevents the pronounced peak of the
relative photon number error exhibited by the fixed stepsize algorithm.

Let us note that an adequate representation of the material dispersion for
simulating the propagation dynamics of ultrashort optical pulses is critical
for obtaining accurate numerical results
\cite{Oughstun:CSE:2003,Oughstun:PRL:1997}.  Thus, for simulations in the
few-cycle regime, a truncated Taylor expansion of the propagation constant,
such as Eq.~(\ref{eq:use_case_2_beta}), in conjunction with a highly accurate
propagation algorithm can be counterproductive.

\subsection{Interaction of four pulses in a ESM fiber \label{sec:use_case_3}}

Finally, we consider a complex scenario, involving the simultaneous propagation
of multiple pulses with distinct center frequencies.
In particular, we consider the medium properties of an ``endlessly single mode''
(ESM) nonlinear photonic crystal fiber \cite{Stone:OE:2007}, which we implement
by a rational Pad\'e-approximant of order $[N=8/M=8]$ for the medium refractive
index in the form
\begin{align}
n(\omega) = 1 + \frac{\sum_{n=0}^{N} p_n \omega^n}{ 1 + \sum_{m=1}^{M} q_m \omega^m }.
                                \label{eq:n_ESM}
\end{align}
The parameter sequences $\{p_n\}_{n=0}^N$ and $\{q_m\}_{m=1}^M$ are detailed in
\ref{sec:prop_const_class}, where a convenience class for handling propagation
constants is introduced.
Representing the medium dispersion as in Eq.~(\ref{eq:n_ESM}) has several
advantages \cite{Amiranashvili:OC:2010,Amiranashvili:OQE:2012}.  It gives a
better approximation of the refractive index than truncating a Taylor expansion
for the detuning $\omega-\omega_0$ for some reference frequency $\omega_0$,
avoids rapid divergence for large frequencies, and, in particular, helps to
avoid unnecessary numerical stiffness when solving nonlinear propagation
equations.

\begin{figure}[t!]
\includegraphics[width=\linewidth]{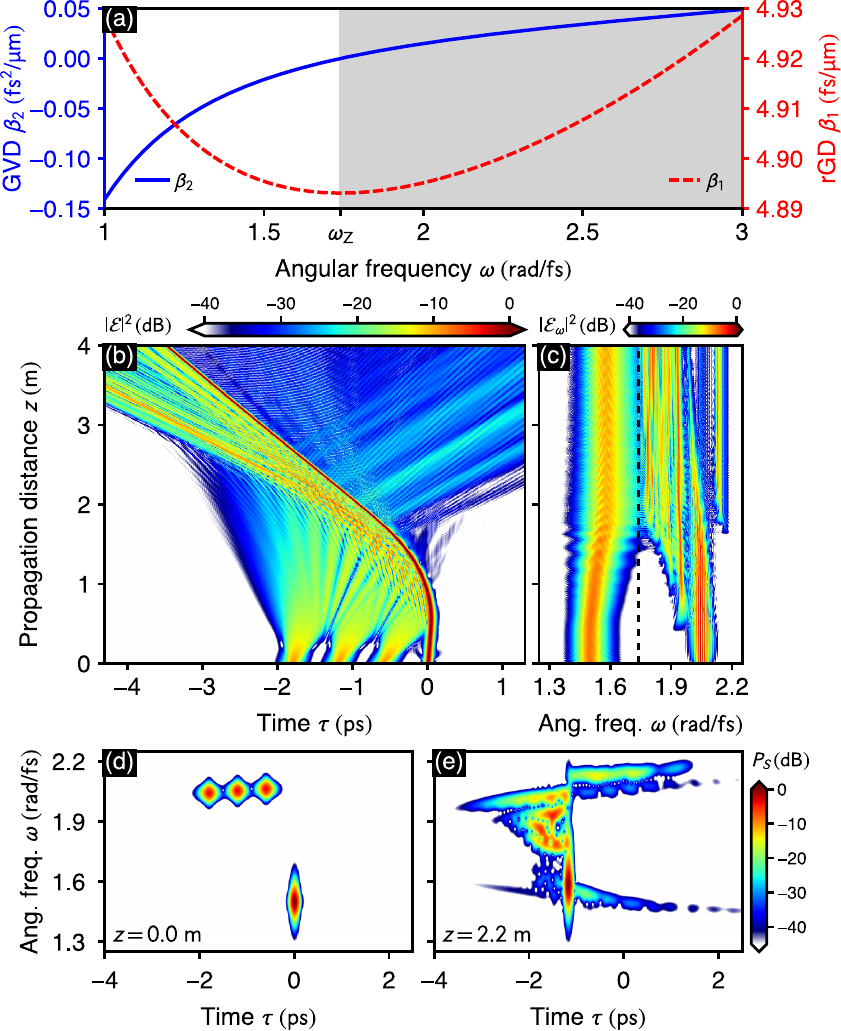}
\caption{
Four pulse interaction in an ESM photonic crystal fiber (PCF).
(a) Frequency dependence of the group-delay (GD) $\beta_1(\omega)$,
and group-velocity dispersion (GVD) $\beta_2(\omega)$.  Shaded region indicates
domain of normal dispersion with zero dispersion point $\omega_{\rm{Z}}\approx
1.741~\mathrm{rad/fs}$. 
Evolution of (b) intensity, and, (c) spectrum over the length of the PCF.
Vertical dashed line in (c) indicates zero-dispersion point.
(d) Spectrogram at $z=0\,\mathrm{m}$, and (e) spectrogram at
$z=2.2\,\mathrm{m}$.
\label{fig:FIG03}}
\end{figure}

\begin{table*}[t!]
\caption{
Simulation parameters used to specify a propagation scenario. These are all
the recognized parameters to be found in the parameter input file. Parameters with 
default setting are optional.
}
\label{tab:sim_pars}
\renewcommand{\arraystretch}{1.2} 
\begin{footnotesize}
\begin{tabular}{lllll}
\hline
\hline
\bf{Parameter} & \bf{Symbol} &  \bf{Value type} & \bf{Description} & \bf{Unit} \\
\hline
{\tt t\_max}    & $t_{\rm{max}}$  
                & {\tt float}         
                & Half-period of temporal mesh        
                & $\mathrm{fs}$
                \\
{\tt t\_num}    & $t_{\rm{num}}$  
                & {\tt int}           
                & Number of mesh points in $t$        
                & --               
                \\
{\tt z\_max}    & $z_{\rm{max}}$  
                & {\tt float}         
                & Value of last mesh-point along $z$  
                & $\mathrm{\mu m}$ 
                \\
{\tt z\_num}    & $z_{\rm{num}}$  
                & {\tt int}           
                & Number of $z$-slices, i.e.\ $z$-steps, along $z$          
                & --               
                \\
{\tt z\_skip}   & $z_{\rm{skip}}$  
                & {\tt int}          
                & Step-interval in which data is stored upon $z$-propagation (default: 1)  
                & --      
                \\
                &                  
                &                      
                & Example: for $z_{\rm{skip}}=4$, data is stored at every 4th integration step
                &
                \\
{\tt beta\_w}   & $\beta(\omega)$  
                & {\tt numpy.ndarray} 
                & Propagation constant     
                & $\mathrm{rad/fs}$      
                \\
{\tt n2}        & $n_2$  
                & {\tt float}         
                & Nonlinear refractive index  
                & $\mathrm{\mu m^2/W}$     
                \\
{\tt fR}        & $f_R$  
                & {\tt float}         
                & Fractional contribution of Raman response 
                & --      
                \\
{\tt tau1}      & $\tau_2$  
                & {\tt float}         
                & Raman response time scale     
                & $\mathrm{fs}$      
                \\
{\tt tau2}      & $\tau_1$  
                & {\tt float}         
                & Raman response time scale     
                & $\mathrm{fs}$      
                \\
{\tt E\_0t}     & $E(0,t)$  
                & {\tt numpy.ndarray} 
                & Real-valued optical field at $z=z_{\rm{min}}$     
                & $\mathrm{\sqrt{W}}$ 
                \\
{\tt out\_file\_path} & -- 
                & {\tt str}      
                & Full path for output file (default: {\tt results.dat})    
                & --      
                \\
\hline
\hline
\end{tabular}
\end{footnotesize}
\end{table*}

The resulting profiles of the group-delay $\beta_1(\omega)$ and group-velocity
dispersion $\beta_2(\omega)$ are shown in Fig.~\ref{fig:FIG03}(a).
For the simulation in terms of the FMAS-S Eq.~(\ref{eq:FMAS_S}) we set
\mbox{$n_2=3\!\cdot\!10^{-8}\,\mathrm{\mu m^2 W^{-1}}$} and neglect the Raman effect.
As initial condition, we consider a fundamental soliton, given by
\begin{align}
E_{\rm{S}}(0,t) = \mathsf{Re}\left[ A_0\, {\rm{sech}}(t/t_{\rm{S}})\,e^{-i\omega_{\rm{S}}t} \right],
\end{align}
with amplitude \mbox{$A_0 =\sqrt{ |\beta_2(\omega_{\rm{S}})| c/(n_2
\omega_{\rm{S}} t_{\rm{S}}^2)}$} and parameters $(t_{\rm{S}}, \omega_{\rm{S}})
= (20\,\mathrm{fs}, 1.5\,\mathrm{rad/fs})$.
We further consider a train of three dispersive waves in the form
\begin{align}
E_{\rm{DW}}(0,t) =\sum_{n=1}^{3}\mathsf{Re}\left[  A_{\rm{DW}}\, {\rm{sech}}\left(\frac{t-\delta_n}{t_{{\rm{DW}}}}\right)\, e^{-i\omega_{n}t}\right],
\end{align}
with common amplitude $A_{\rm{DW}}=0.35\,A_0$, common duration $t_{\rm{DW}}=60\,\mathrm{fs}$,
and parameters
$(\delta_{1},\omega_{1}) = (-0.6\,\mathrm{ps} , 2.06\,\mathrm{rad/fs})$,
$(\delta_{2},\omega_{2}) = (-1.2\,\mathrm{ps} , 2.05\,\mathrm{rad/fs})$, and
$(\delta_{3},\omega_{3}) = (-1.8\,\mathrm{ps} , 2.04\,\mathrm{rad/fs})$.
%
For the computational domain we choose $t_{\rm{max}}=8\,\mathrm{ps}$,
$t_{\rm{num}}=2^{15}$, and propagate up to $z_{\rm{max}}= 6\,\mathrm{m}$ using
step size $\Delta z=80\,\mathrm{\mu m}$, For $z$-propagation we use the
IFM-RK4IP method.
The propagation dynamics of the field in both, time domain and frequency
domain, is detailed in Figs.~\ref{fig:FIG03}(b-c).  
Specifically, Fig.~\ref{fig:FIG03}(b) shows the time-domain intensity of the analytic
signal for the retarded time coordinate $\tau = t - z/v_0$, with reference
velocity $v_0 = 1/\beta_1(\omega_{\rm{S}})$.
As evident from the spectrogram (see \ref{sec:spectrograms}) in
Fig.~\ref{fig:FIG03}(d), all four pulses can be distinguished very well for the
initial condition at $z=0\,\mathrm{m}$.  Upon propagation, a complex dynamics
unfolds as can be seen from Figs.~\ref{fig:FIG03}(b-c) and the spectrogram at
$z=2.2\,\mathrm{m}$ [Fig.~\ref{fig:FIG03}(e)].  
Therein, the soliton induces a strong refractive index barrier for the
dispersive waves in the domain of normal dispersion \cite{Demircan:PRL:2011},
leading to multiple scattering processes.
The underlying process is enabled by a general wave reflection mechanism,
originally reported in fluid dynamics \cite{Smith:PCPS:1975}.  In optics it is
referred to as the push-broom effect \cite{deSterke:OL:1992}, optical event
horizon \cite{Philbin:Science:2008,Faccio:CP:2012}, or temporal reflection
\cite{Plasinis:PRL:2015}.  It allows for a strong and efficient all optical
control of light by light \cite{Demircan:PRL:2013, Demircan:OL:2014}, and  has
been shown to naturally appear in the process of supercontinuum generation
\cite{Driben:OE:2010,Demircan:SR:2012,Demircan:APB:2014, Armaroli:O:2015}.

The above propagation scenario illustrates the simulation of complex
short-pulse interaction dynamics, as, e.g., given in all-optical supercontinuum
switching \cite{Melchert:NC:2020}. The simulations in
Ref.~\cite{Melchert:NC:2020} where performed using the {\tt py-fmas} library
code.

\appendix

\section{Recognized input-file parameters\label{sec:input}}

As discussed in sect.~\ref{sec:workflow_a}, {\tt py-fmas} provides convenience
methods that read a user-defined propagation scenario from an adequate input
file in HDF5 format.  In that case, the input file must contain all parameters
needed to specify the computational domain, propagation model, and propagation
algorithm.  In Tab.~\ref{tab:sim_pars} we list the recognized parameters that
can be supplied in terms of such an input file.  All parameters without default
values must be present.

\section{Computing spectrograms\label{sec:spectrograms}}

A spectrogram provides a particular time-frequency representation of a
considered signal and represents an integral tool in the analysis and characterization
of ultrashort optical pulses, both in theory
\cite{Dudley:OE:2002,Skryabin:PRE:2005} and experiment
\cite{Trebino:JQE:1993,Linden:BOOK:2000,Efimov:PRL:2005}.
{\tt py-fmas} includes the functionality to compute a spectrogram of the
time-domain analytic signal $\Ecal(z,t)$ at a given $z$-coordinate in terms of
a short-time Fourier transform as
\begin{align}
P_{S}(t,\omega) = \frac{1}{2 \pi} \left|\int \Ecal(z,t^\prime)h(t^\prime-t) e^{-i \omega t}~{\rm d}t^\prime\right|^2, \label{eq:PS}
\end{align}
wherein $h(x)=\exp(-x^2/2\sigma^2)$ specifies a Gaussian window function with
root-mean-square width $\sigma$, used to localize $\Ecal(z,t)$ in time.
For computing such spectrograms, module {\tt tools} provides the function 
\begin{verbatim}
spectrogram(t, w, ut, t_lim, Nt, Nw, s0)
\end{verbatim}
where {\tt t} (type {\tt numpy.ndarray}) is the $t$-grid Eq.~(\ref{eq:grid_t}),
{\tt w} (type {\tt numpy.ndarray}) is the $\omega$-grid Eq.~(\ref{eq:grid_w}),
{\tt ut} (type {\tt numpy.ndarray}) is the analytic signal $\Ecal(z,t)$ at a
given $z$-coordinate,
{\tt t\_lim} (type {\tt list}) is a 2-tuple specifying bounds for the $t$-axis
when computing the spectrogram (default: $(\min(t),\max(t))$),
{\tt Nt} (type {\tt int}) is the number of equidistant samples used for
localization along $t$ (default: $1000$),
{\tt Nw} (type {\tt int}) is the number of equidistant $\omega$-samples kept on
output (default: $2^8$),
and {\tt s0} (type {\tt float}) is the RMS width of $h$ in Eq.~(\ref{eq:PS})
(default: $20\,\mathrm{fs}$).  
Upon termination, the above function returns the
3-tuple $({\tt tS}, {\tt wS}, {\tt PS})$, with {\tt tS}  (type {\tt
numpy.ndarray}), and {\tt wS}  (type {\tt numpy.ndarray}) the spectrograms
discrete $t$ and $\omega$ axes, and {\tt PS} (type {\tt numpy.ndarray}) the
corresponding two-dimensional spectrogram trace.
A function with call-signature {\tt plot\_spectrogram(tS, wS, PS)}, assisting a
user to quickly visualize the spectrogram data, is also included in module {\tt
tools}.

Note that {\tt py-fmas} can also be used in conjunction with the {\tt optfrog}
spectrogram tool \cite{Melchert:SFX:2019}, allowing a user to calculate
spectrograms with optimized time-frequency resolution.  Examples that
illustrate how to use the above functions as well as how to blend {\tt py-fmas}
with {\tt optfrog} are provided along with the \href{\LINKTOCODE}{online
documentation}.

\section{Raman response functions\label{sec:hR}}

Numerical models of the Raman response are important for the accurate
theoretical description of the propagation of optical pulses with short
duration and high peak power \cite{Mitschke:OL:1986,Gordon:OL:1986}.  For
example, the Raman response includes the self-frequency shift that affects the
propagation dynamics of solitons.
{\tt py-fmas} implements several models of the Raman response function, located
in module {\tt raman\_response}.  Specifically, the implemented models are:

\begin{itemize}
\item 
{Blow-Wood type response function \cite{Blow:JQE:1989}:}
The time-domain formulation of this response function, based on a
single-damped-harmonic-oscillator approximation with Lorentzian linewidth,
reads
\begin{equation}
h_{\mathrm{BW}}(t) = \frac{\tau_1^2 + \tau_2^2}{\tau_1\tau_2^2}\, e^{-t/\tau_2}\, \sin(t/\tau_1)\,\theta(t), \label{eq:RamanBW}
\end{equation}
where causality is assured by the unit step function $\theta(t)$.  
Equation~(\ref{eq:RamanBW}) represents a generic two parameter response function
that can be adapted to fit various types of nonlinear fibers. For example,
using a fractional Raman contribution $f_R=0.18$ [cf.\ Eq.~(\ref{eq:FMAS_S_R})]
together with $\tau_1=12.2\,\mathrm{fs}$, and $\tau_2=32\,\mathrm{fs}$ is
adequate for simulation of silica fibers \cite{Blow:JQE:1989}.
Using $f_R=0.1929$, $\tau_1=9\,\mathrm{fs}$, and $\tau_2=134\,\mathrm{fs}$ is
adequate for ZBLAN fluoride fibers \cite{Liu:OE:2011,Agger:JOSAB:2012}.
This response model is implemented as function
\mbox{{\tt h\_BW(t, tau1, tau2)}}, where {\tt t} (type {\tt numpy.ndarray}) is
the $t$-grid, and {\tt tau1} (type {\tt float}, default: $12.2\,\mathrm{fs}$),
and {\tt tau2} (type {\tt float}, default: $32.\,\mathrm{fs}$) are the two
parameters with default values valid for fused silica.

As detailed in Eq.~(\ref{eq:I_Raman}), by default, our propagation model
\mbox{FMAS-S-R} implements the corresponding frequency-domain representation.
However, for completeness, we also provide the implementation according to
Eq.~(\ref{eq:RamanBW}).

\begin{table}[b!]
\renewcommand{\arraystretch}{1.} 
\caption{
Parameters defining the Raman response function $h_{\mathrm{HC}}$ [see
Eq.~(\ref{eq:RamanHC})].
From left to right: number of vibrational mode ($n$), vibrational frequency
($\omega_n$), mode amplitude ($A_n$), Lorentzian linewidth ($\gamma_n$), and
Gaussian linewidth ($\Gamma_n$).
Values are taken from Ref.~\cite{Hollenbeck:JOSA:2002}.
}\label{tab:RamanHC}
\begin{footnotesize}
\begin{tabular}{r S[table-format=1.5] S[table-format=3.2] S[table-format=2.2]  S[table-format=3.2] }
\hline
\hline
$n$ & $\omega_n~\mathrm{\left(\frac{rad}{fs}\right)}$ & $A_n\,\mathrm{(-)}$ & $\gamma_n\,\mathrm{(\times 10^{-3}\,fs^{-1})}$ & $\Gamma_n\,\mathrm{(\times 10^{-3}\,fs^{-1})}$\\
\hline
1 & 0.01060 &  1.00 &  1.64 &  4.91 \\
2 & 0.01884 & 11.40 &  3.66 & 10.40 \\
3 & 0.04356 & 36.67 &  5.49 & 16.48 \\
4 & 0.06828 & 67.67 &  5.10 & 15.30 \\
5 & 0.08721 & 74.00 &  4.25 & 12.75 \\
6 & 0.09362 &  4.50 &  0.77 &  2.31 \\
7 & 0.11518 &  6.80 &  1.30 &  3.91 \\
8 & 0.13029 &  4.60 &  4.87 & 14.60 \\
9 & 0.14950 &  4.20 &  1.87 &  5.60 \\
10 & 0.15728 &  4.50 &  2.02 &  6.06 \\
11 & 0.17518 &  2.70 &  4.71 & 14.13 \\
12 & 0.20343 &  3.10 &  2.86 &  8.57 \\
13 & 0.22886 &  3.00 &  5.02 & 15.07 \\ 
\hline
\hline
\end{tabular}
\end{footnotesize}
\end{table}

\item {Lin-Agrawal type response function \cite{Lin:OL:2006}:}
The time-domain formulation of this response function, based on an improved
model that takes into account the anisotropic nature of Raman scattering, reads
\begin{align}
h_{\mathrm{LA}}(t) = (1-f_b)\,h_{\mathrm{BW}}(t) +f_b\,\frac{2\tau_b-t}{\tau_b^2} e^{-t/\tau_b}\,\theta(t), \label{eq:RamanLA}
\end{align}
with $h_{\mathrm{BW}}(t)$ [see Eq.~(\ref{eq:RamanBW})] modeling the isotropic
part of the response, $\tau_b=96\,\mathrm{fs}$, and $f_b = 0.21$. 
This response model is implemented as function
\mbox{{\tt h\_LA(t)}}, where {\tt t} (type {\tt numpy.ndarray}) is
the $t$-grid. 

\item {Hollenbeck-Cantrell type response function \cite{Hollenbeck:JOSA:2002}:}
This elaborate response function implements the intermediate broadening model
for the Raman response of silica fibers detailed in
Ref.~\cite{Hollenbeck:JOSA:2002}. The time-domain representation of this model
reads
\begin{align}
h_{\mathrm{HC}}(t) = \sum_{n=1}^{13} A_n\, e^{-\gamma_n t - \Gamma_n^2 t^2/4}\,\sin(\omega_n t)\,\theta(t), \label{eq:RamanHC}
\end{align}
with parameter sequences $\{\omega_n\}_{n=1}^{13}$, $\{A_n\}_{n=1}^{13}$, $\{\gamma_n\}_{n=1}^{13}$, and $\{\Gamma_n\}_{n=1}^{13}$, summarized in Tab.\ \ref{tab:RamanHC}.
In Eq.~(\ref{eq:RamanHC}), each term represents a specific vibrational mode of
$\mathrm{Si_2O}$. 
The parameters in Tab.~\ref{tab:RamanHC} are derived from spectroscopic 
data obtained for a fused silica fiber \cite{Stolen:JOSA:1989}.
This response model is implemented as function
\mbox{{\tt h\_HC(t)}}, where {\tt t} (type {\tt numpy.ndarray}) is
the $t$-grid. 
\end{itemize}

The response functions defined by Eqs.~(\ref{eq:RamanBW},\ref{eq:RamanLA}) are
quite generic.  In contrast, the Hollenbeck-Cantrell type Raman model
Eq.~(\ref{eq:RamanHC}) is rather specific and caution is needed to ensure it is
not used out of its range of applicability.
An example that shows how the above Raman response functions can be used with
the models implemented by {\tt py-fmas} is provided along with the
\href{\LINKTOCODE}{online documentation}.

\begin{lstlisting}[
    float,
    floatplacement = b!, 
    numbers = left,
    captionpos = t,
    keywordstyle = \bf,
    frame = lines,
    stepnumber = 1,
    numbers = left, 
    numbersep = 5pt, 
    xleftmargin = \parindent,
    language = Python,
    basicstyle = \ttfamily\scriptsize, 
    caption = {Script demonstrating how the convenience class {\tt PropConst}
    can be used to analyze a propagation constant.}, 
    label = code:pc]
import numpy as np
from fmas.propagation_constant import PropConst

def get_beta_fun_ESM():
    p = np.poly1d((16.89475, 0, -319.13216, 0,
      34.82210, 0, -0.992495, 0, 0.0010671)[::-1])
    q = np.poly1d((1.00000, 0, -702.70157, 0,
      78.28249, 0, -2.337086, 0, 0.0062267)[::-1])
    c = 0.29979 # (micron/fs)
    return lambda w: (1+p(w)/q(w))*w/c

beta_fun = get_beta_fun_ESM()
pc = PropConst(beta_fun)

w_Z = pc.find_root_beta2(1.3, 2.2)
# -- YIELDS: w_Z = 1.740823 rad/fs

w_S = 1.5 # (rad/fs)
w_GVM = pc.find_match_beta1(w_S, w_Z, 2.5)
# -- YIELDS: w_GVM = 2.019102 rad/fs 

w_DW1 = 2.06 # (rad/fs)
dvg = pc.vg(w_DW1) - pc.vg(w_S)
# -- YIELDS: dvg = -0.000029 micron/fs

betas = pc.local_coeffs(w_S, n_max = 4)
# -- YIELDS: betas =
# [ 7.220, 4.8954, -0.0105, 0.0184, -0.0103]
#  fs/mu,  fs^2/mu, fs^3/mu, fs^4/mu, fs^5/mu; mu=micron
\end{lstlisting}

\section{Analyzing propagation constants \label{sec:prop_const_class}}

The design of custom propagation scenarios that either match experiments or
carve out specific effects, observed during the dynamical evolution of optical
pulses, typically require the analysis of a propagation constant.
To assist a user in doing this, {\tt py-fmas} provides the convenience class
{\tt PropConst}, defined in module {\tt propagation\_constant}, allowing to
wrap and analyze a user defined propagation constant $\beta(\omega)$. A
prerequisite for using {\tt PropConst} is that $\beta(\omega)$ needs to be
available as callable function.

A basic example illustrating some of the functionality implemented in terms of 
{\tt PropConst} is shown in listing \ref{code:pc}.
Therein, in lines 4--10, the propagation constant $\beta(\omega) =
\omega\,n(\omega)/c$ for an ``endlessly single mode'' (ESM) photonic crystal
fiber \cite{Stone:OE:2007} is defined.
The enclosing function {\tt get\_beta\_fun\_ESM} returns a closure,
implementing a rational Pad\'e-approximant of order $[N\!=\!8/M\!=\!8]$ for the
refractive index $n(\omega)$, as defined by Eq.~(\ref{eq:n_ESM}).
In line 12, $\beta(\omega)$ is initialized. It is
wrapped by the convenience class in line 13.
In line 15 it is shown how a zero-dispersion point, located within the
bracketing interval $(1.3, 2.2)\,\mathrm{rad/fs}$ can be determined, yielding
$\omega_{\rm{z}}\approx 1.7408\,\mathrm{rad/fs}$.
In line 19 it is shown how a group-velocity matched (GVM) partner frequency to
$\omega_{\mathrm{S}}=1.5\,\mathrm{rad/fs}$, located in the bracketing interval
$(\omega_{\rm{Z}},2.5 \,\mathrm{rad/fs})$, can be determined, giving 
$\omega_{\rm{GVM}}\approx2.019\,\mathrm{rad/fs}$.
The group-velocity mismatch $\Delta v_g =
v_g(\omega_{\rm{DW1}})-v_g(\omega_{\rm{S}})$ for
$\omega_{\rm{DW1}}=2.06\,\mathrm{rad/fs}$ [see sect.~\ref{sec:use_case_3}] is
calculated in line 23, yielding $\Delta v_g \approx
-2.9\cdot10^{-5}\,\mathrm{\mu m/fs}$.
Finally, in line 26 it is shown how the Taylor-expansion coefficients
$\{\beta_n\}_{n=0}^4$ at $\omega_{\rm{S}}$ can be obtained, yielding 
$\beta_0\approx 7.220\,\mathrm{\mu m^{-1}}$, 
$\beta_1\approx 4.8954\,\mathrm{fs/\mu m}$, 
$\beta_2\approx -0.0105\,\mathrm{fs^2/\mu m}$, 
$\beta_3\approx 0.0184\,\mathrm{fs^3/\mu m}$, and 
$\beta_4\approx -0.0103\,\mathrm{fs^4/\mu m}$.






\addcontentsline{toc}{section}{Acknowledgements}

\section*{Acknowledgements}

We acknowledge financial support from the Deutsche Forschungsgemeinschaft
(DFG) under Germany’s Excellence Strategy within the Clusters of Excellence
PhoenixD (Photonics, Optics, and Engineering – Innovation Across Disciplines)
(EXC 2122, projectID 390833453).

\end{document}